\documentclass{aa}
\usepackage{graphicx}

\def\jh{\mbox{$\rm (J-H)$}}

\def\mM{\mbox{$\rm (m-M)_0$}}
\def\ebv{\mbox{$\rm E(B-V)$}}
\def\ejh{\mbox{$\rm E(J-H)$}}

\def\rc{\mbox{$\rm R_{core}$}}

\def\rlim{\mbox{$\rm R_{lim}$}}
\def\ms{\mbox{$\rm M_\odot$}}
\def\ds{\mbox{$\rm d_\odot$}}
\def\mj{\mbox{$\rm M_J$}}
\def\jj{\mbox{$\rm J$}}
\def\hh{\mbox{$\rm H$}}
\def\mg{\mbox{$\rm m_{giant}$}}
\def\mMs{\mbox{$\rm m_{MS}$}}
\def\mobs{\mbox{$\rm m_{obs}$}}
\def\age{\mbox{$\tau_{age}$}}
\def\zz{\mbox{$[Z/Z_\odot]$}}

\begin{document}

\title{NGC\,2180: a disrupting open cluster}

\author{C. Bonatto \inst{1}, E. Bica \inst{1} \and D. B. Pavani\inst{1}}

\offprints{Ch. Bonatto - charles@if.ufrgs.br}

\institute{Universidade Federal do Rio Grande do Sul, Instituto de F\'\i sica, 
CP\,15051, Porto Alegre 91501-970, RS, Brazil\\
\mail{}
}

\date{Received --; accepted --}

\abstract{The spatial dependence of luminosity and mass functions of evolved open clusters 
is discussed in this work using J and H 2MASS photometry, which allows a wide spatial 
coverage and proper background determination. The target objects are the overlooked open 
cluster NGC\,2180 ($\ell=203.85^\circ$, $b=-7.01^\circ$) and the intermediate-age open 
cluster NGC\,3680 ($\ell=286.76^\circ$\ and $b=16.92^\circ$), which has been reported as 
being in an advanced state of dissolution. We conclude that, although in an advanced 
dynamical state (mass segregated), NGC\,3680 does not present strong signs of dissolution, 
having luminosity and mass functions very similar to those of the $\age\approx3.2$\,Gyr 
open cluster M\,67. On the other hand, NGC\,2180 presents flat, eroded luminosity functions 
throughout its structure, indicating that in addition to mass segregation, Galactic tidal 
stripping has been effective in depleting this cluster of stars. Accordingly, NGC\,2180 may 
be the missing link between evolved open clusters and remnants. For NGC\,2180 we derive an 
age $\age\approx710$\,Myr and an observed stellar mass of $\mobs\sim47\ms$. Most of the 
colour-magnitude diagram features,  the main sequence in particular, are equally 
well reproduced by isochrones with metallicity $\zz=-0.4$ and 0.0. The solution 
for $\zz=-0.4$ results in $\mM=9.40\pm0.20$, $\ebv=0.18$ and a distance to the Sun 
$\ds=0.76\pm0.06$\,kpc, while that for $\zz=0.0$ gives $\mM=10.10\pm0.20$, $\ebv=0.0$ and 
$\ds=1.05\pm0.08$\,kpc. For NGC\,3680 we derive an age $\age\approx1.6$\,Gyr, $\ebv=0.0$ 
and $\ds=1.00\pm0.09$\,kpc, confirming previous estimates. The observed stellar mass 
$\mobs\approx130\,\ms$ agrees with previous values. We study both clusters in the context 
of dynamical states estimated from diagnostic-diagrams involving photometric and structural 
parameters. Both clusters are dynamically evolved systems. In particular, NGC\,2180 is 
closer to open cluster remnants than NGC\,3680.

 }

\titlerunning{Open cluster evolution}

\authorrunning{C. Bonatto et al.}

\maketitle

\section{Introduction}
\label{intro}

Open clusters are formed along the gas and dust-rich Galactic plane and
contain from tens to a few thousands of stars distributed in an approximately 
spherical structure of up to a few parsecs in radius. This loose condition makes 
them potentially short-lived stellar systems, since disruptions may occur by the 
cumulative effect of passages near interstellar clouds and/or by shocks with the 
Galactic disk. Cumulative orbital perturbations may lead to more internal orbits, 
enhancing such effects Bergond et al. (\cite{Bergond2001}). Consequently, most of 
the open clusters in the Galaxy evaporate completely in less than 1\,Gyr. Indeed, 
the open cluster catalogue of Lyng\aa\ (\cite{Lyngaa1987}) indicated about 70 objects 
older than 1\,Gyr ($\approx6\%$ of the total number).

The dynamical evolution of an open cluster depends both on internal and external factors.
Internal factors are: {\it (i)} after successive 2-body encounters with more massive 
stars, less-massive stars may acquire velocities larger than the cluster's escape velocity, 
and {\it (ii)} the normal stellar evolution via mass-loss. The external factors are:
{\it (i)} large-scale encounters with giant molecular clouds (Wielen \cite{Wielen1991}), 
and {\it (ii)} tidal stripping by the Galactic gravitational field. A typical open cluster 
at the solar radius will cross the Galactic plane 10--20 times before being disrupted and 
leaving an open cluster remnant (de la Fuente Marcos \cite{delaF1998}). Bergond et al.
(\cite{Bergond2001}) estimate the destruction time-scale for open clusters
in the solar neighbourhood at about 600\,Myr. Consequently, it is expected that
only those open clusters which are born with the largest masses or those located at
large Galactic radii will survive longer than a few Gyr (Friel \cite{Friel1995}).

The numerical simulations of de la Fuente Marcos (\cite{delaF1996}) have shown that
the final cluster remnant composition depends on the initial mass function, fraction 
of primordial binaries and galactocentric distance. The resulting cluster remnants 
are rich in binaries and do not appear to contain collapsed objects. Remnants of poorly 
populated clusters are expected to contain early-type stars, while those of more massive
clusters contain late-type stars (de la Fuente Marcos \cite{delaF1996}), owing to
different evolutionary time-scales. In the central region of the more evolved clusters, 
mass segregation should deplete the low main-sequence (MS) stars thus creating a core 
rich in compact and giant stars (Takahashi \& Portegies Zwart \cite{TakaP2000}).

Mass segregation in a star cluster scales with the relaxation time, defined as
$t_{relax}=\frac{N}{8\ln N}t_{cross}$, where $t_{cross}=R/v$ is the crossing
time (Binney \& Tremaine \cite{BinTre1987}). For a typical cluster radius of $R\sim5$\,pc 
and velocity dispersion $v\sim1$\,km\,s$^{-1}$, $t_{relax}\sim13$\,Myr
for a cluster with $N=10^2$ stars, and $t_{relax}\sim90$\,Myr for $N=10^3$ stars.

Recently, several studies called attention to the possibility of detecting open cluster
remnants in the Galaxy, e.g. Bica et al. (\cite{Bica2001}), Carraro (\cite{Carraro2002}), 
Pavani et al. (\cite{Pavani2002}, \cite{Pavani2003}). A fundamental question to dynamical 
evolution studies is whether any open cluster can be observed right at the disrupting phase, 
when the remaining low-mass stars in the cluster's halo get dispersed into the background 
and the corresponding mass function becomes eroded.

Depletion of low-MS stars in the central parts of a cluster is a sign of
advanced dynamical evolution. This has been detected e.g. in NGC\,3680 
(Anthony-Twarog et al. \cite{Twa1991}) and M\,67 (Bonatto \& Bica \cite{BB2003}),
in which the presence of a corona rich in low-mass stars has been confirmed
with 2MASS photometry. 

The Two Micron All Sky Survey (hereafter 2MASS, Skrutskie et al. \cite{2mass1997}) 
has proven to be a powerful tool in the analyses of the structure and stellar content 
of open clusters (e.g. Bonatto \& Bica \cite{BB2003}, Bica et al. \cite{BBD2003}).
Indeed, the uniform and essentially complete sky coverage provided by 2MASS allows
one to properly take into account background regions with suitable star count
statistics, which is fundamental in order to correctly identify and characterize
the stellar content of clusters, since their ages and distances can be determined 
by fitting isochrones to their colour-magnitude diagrams (CMDs), with a precision 
depending on the depth of the photometry and field contamination. 

In the present study we address the actual dynamical state of NGC\,3680 analysing a 
large spatial area in the direction of the cluster, which 2MASS can provide. In 
particular, we search for the presence of a low-mass star-rich corona. In addition, 
we discuss NGC\,2180, an overlooked open cluster. This cluster appears to be in a 
more advanced dynamical evolutionary stage than NGC\,3680, and thus might be a missing 
link between evolved open clusters with a corona and final remnants.

In Section~\ref{targets} we provide available information on NGC\,2180 and the
intermediate-age open cluster NGC\,3680. In Sect.~\ref{2massPh} we obtain the 2MASS 
photometry and introduce the $\jj\times\jh$ CMDs. In Sect.~\ref{StructAnal} we discuss 
the radial density distribution of stars and derive structural parameters for the 
clusters. In Sect.~\ref{Fund_par} we fit isochrones to the near-infrared CMDs and 
derive cluster parameters. In Sect.~\ref{LumFunc} we derive the luminosity and mass 
functions (hereafter LF and MF) and estimate the stellar masses of each cluster.  
In Sect.~\ref{Comp} we compare NGC\,2180 and NGC\,3680 with well-known dynamically 
evolved open clusters and open cluster remnants. Concluding remarks are given in 
Sect.~\ref{Conclu}. 

\section{The target clusters}
\label{targets}

\subsection{NGC\,2180: a W. Herschell's overlooked open cluster}
\label{N2180}

NGC\,2180 has been first observed and described as a cluster by W. Herschell (Dreyer 
\cite{Dreyer1888}). In modern catalogues of open clusters, it is not included (Alter et 
al. \cite{Alter1970}; Lyng\aa~\cite{Lyngaa1987}). In Dias et al. (\cite{Dias2002}) and 
in a revision of the NGC catalogue (Sulentic \& Tifft \cite{Sulentic1973}), NGC\,2180
is indicated as non-existing. Houston (\cite{Houston1976}) included it in a list of possible 
clusters. The only information currently available in the WEBDA open cluster database 
(Mermilliod \cite{Merm1996} --- {\em http://obswww.unige.ch/webda} is the object's 
designation.

\begin{figure}
\resizebox{\hsize}{!}{\includegraphics{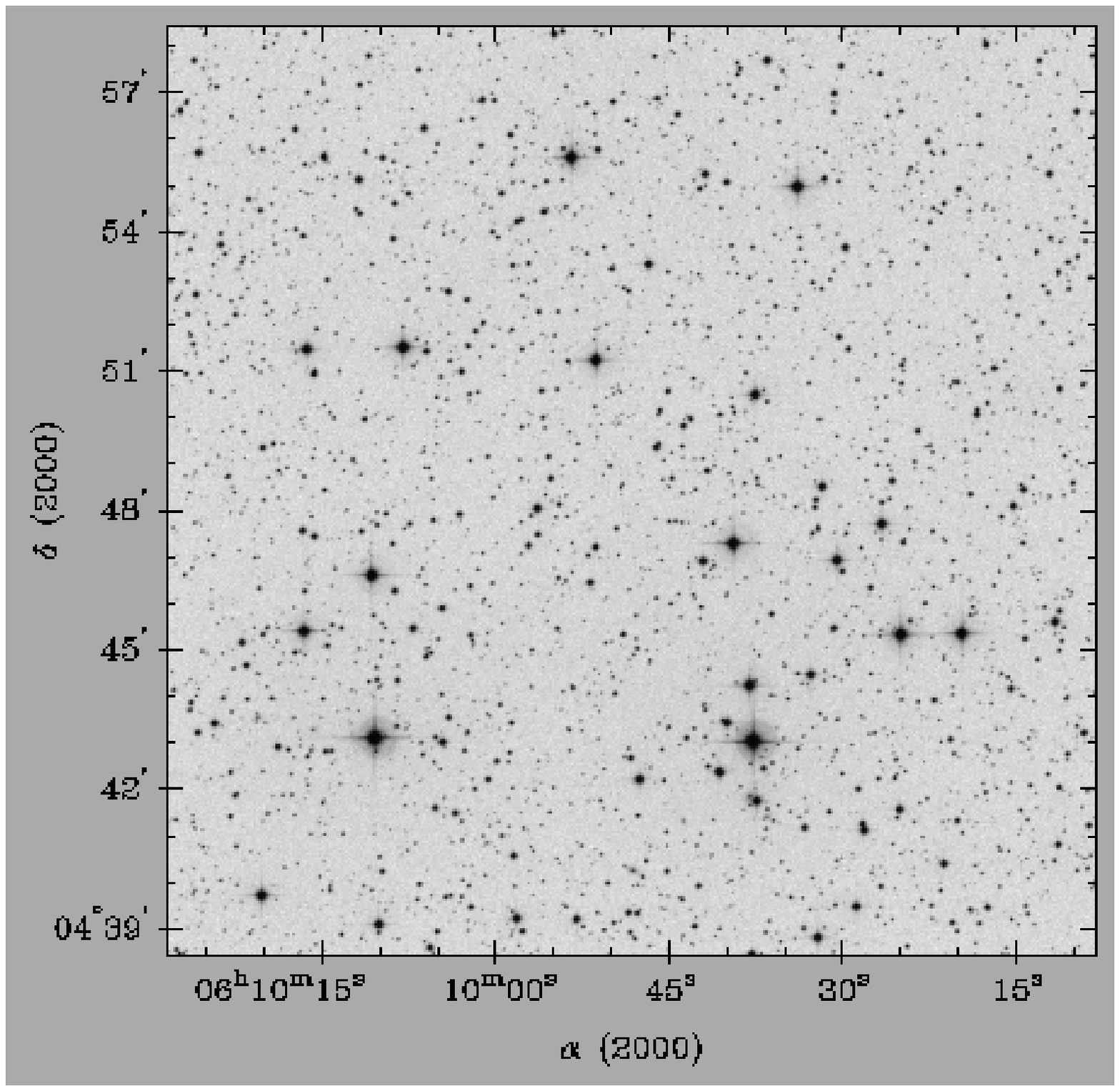}}
\caption[]{$20\arcmin\times20\arcmin$  ($\approx5.3\times5.3\rm\,pc^2$) XDSS R 
image of NGC\,2180.}
\label{fig1}
\end{figure}

The original coordinates of NGC\,2180 precessed to J2000 are 
$\alpha=06^h\,09^m\,36^s$\ and $\delta=+04^\circ\,43\arcmin$. However,
in what follows we revised the central coordinates to $\alpha=06^h\,09^m\,48^s$\
and $\delta=+04^\circ\,48\arcmin\,26\arcsec$ based on bright star membership
(Sect.~\ref{2massPh}). The latter coordinates convert to $\ell=203.85^\circ$\
and $b=-7.01^\circ$. A Digitized Sky Survey (XDSS) R image of NGC\,2180,
centered on the revised coordinates is given in Fig.~\ref{fig1}.

Visually, the cluster is sparse and poorly populated, but a few bright stars
are present in the area.

\subsection{The dynamically evolved open cluster NGC\,3680}
\label{N3680}

The intermediate-age open cluster NGC\,3680 has already been extensively
studied in the past. Its slightly  supersolar metallicity has been determined
by Eggen (\cite{Eggen1969}) and through the photoelectric {\it uvby}-H$_\beta$
photometry of Nissen (\cite{Nissen1988}). CCD {\it uvby} photometry of the
central parts of NGC\,3680 has been published by Anthony-Twarog et al.
(\cite{Twa1989}), and CCD and photographic BV photometry by Anthony-Twarog et al.
(\cite{Twa1991}).

Previous age estimates for NGC\,3680 varied from $\age\sim1.0$\,Gyr (Mazzei \& Pigatto
\cite{MP1988}) to $\age\sim4.5$\,Gyr (Anthony-Twarog et al. \cite{Twa1991}). The
Str\"omgren photometry of Bruntt et al. (\cite{Bruntt1999}) resulted in an age of
$\age=1.45\pm0.15$\,Gyr, comparable to the value $\age=1.6\pm0.5$\,Gyr derived
by Kozhurina-Platais et al. (\cite{Kozhu1997}).

With respect to the dynamical state, Nordstr\"om et al. (\cite{Nordstrom1996}),
based on {\it by} CCD photometry of 310 stars as well as radial velocity and
proper motion data, concluded that NGC\,3680 should be the last remnant of a 
cluster in an advanced state of dissolution, almost lost in the foreground field 
of similar stars. They estimate the present stellar mass as $\mobs\sim100\,\ms$,
and an initial total mass of $\sim1200\,\ms$.

\begin{figure}
\resizebox{\hsize}{!}{\includegraphics{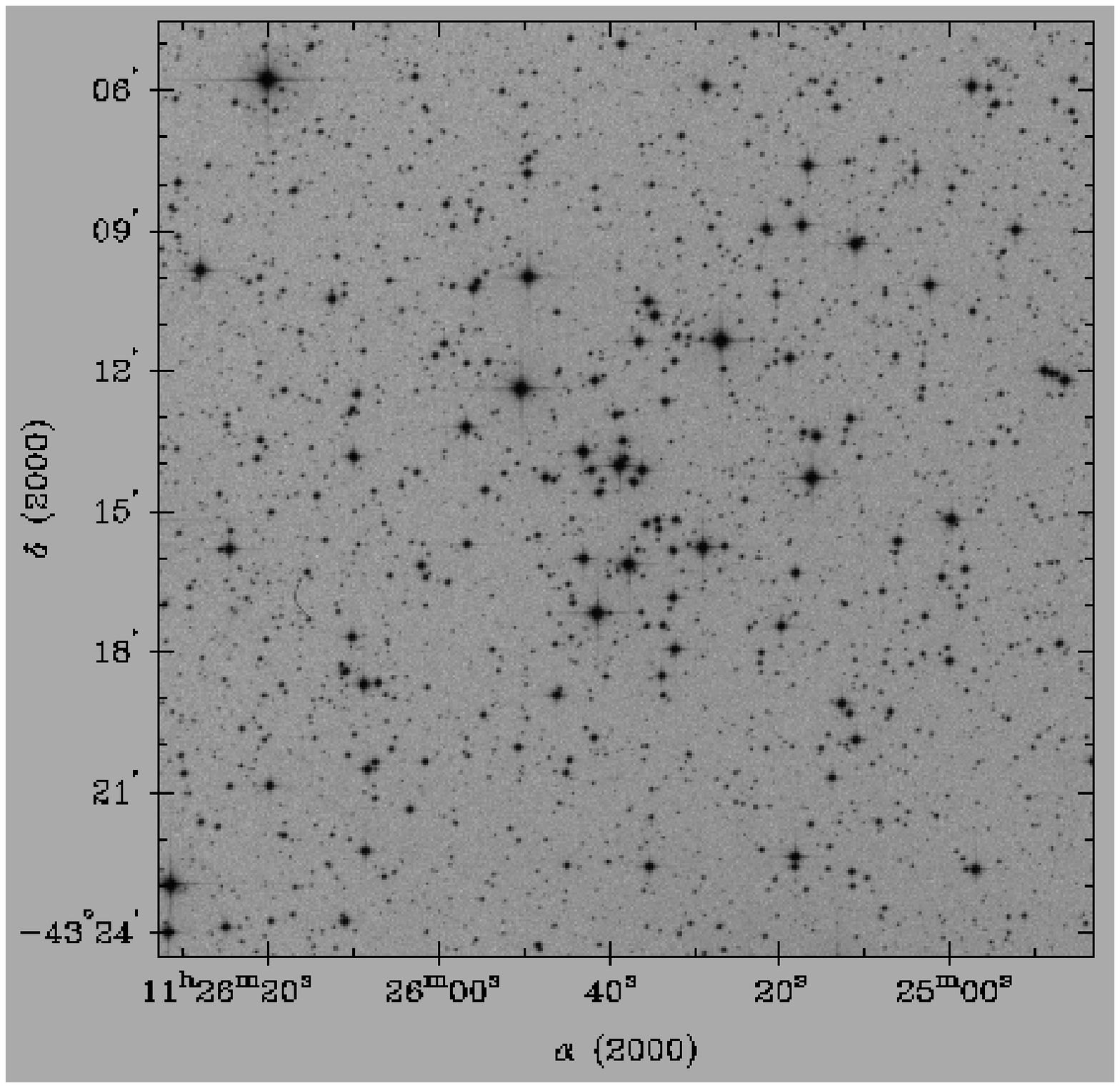}}
\caption[]{$20\arcmin\times20\arcmin$  ($\approx5.8\times5.8\rm\,pc^2$) XDSS R image 
of NGC\,3680.}
\label{fig2}
\end{figure}

The central coordinates of this intermediate-age open cluster, estimated from the 
XDSS image, are $\alpha=11^h\,25^m\,38^s$\ and $\delta=-43^\circ\,14\arcmin\,30\arcsec$. 
These values agree with the WEBDA coordinates. The new coordinates convert to
$\ell=286.76^\circ$\ and $b=16.92^\circ$. A Digitized Sky Survey (XDSS) R image of 
NGC\,3680 is given in Fig.~\ref{fig2}, in which a concentration of bright stars is
present in the central $8\arcmin\times8\arcmin$.

\section{The 2MASS photometry}
\label{2massPh}

We investigate the nature and structure of both clusters using J and H photometry
obtained from the 2MASS All Sky data release, which is available at
{\em http://www.ipac.caltech.edu/2mass/releases/allsky/}. 2MASS photometric errors
typically attain 0.10\,mag at $\jj\approx16.2$ and $\hh\approx15.0$, see e.g. Soares 
\& Bica (\cite{SB2002}). Star extractions have been performed using the VizieR tool 
at {\em http://vizier.u-strasbg.fr/viz-bin/VizieR?-source=2MASS}. For each cluster we 
made circular extractions centered on the coordinates given in Sect.~\ref{targets}. 
In order to maximize the
statistical significance and representativity of background star counts, we decided
to use an external ring (same area as the cluster) as offset field. This offset field
will be used to represent the stellar background contribution to the cluster.
We used an extraction radius of 40\arcmin\ for NGC\,2180 and 30\arcmin\ for NGC\,3680.

\begin{figure}
\resizebox{\hsize}{!}{\includegraphics{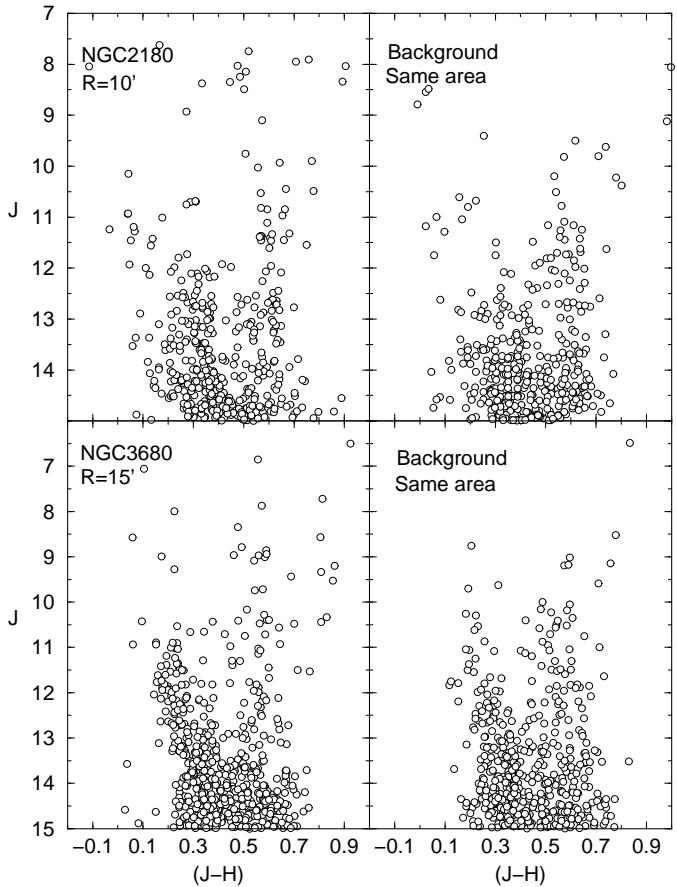}}
\caption[]{$\jj\times\jh$ CMDs for the cluster NGC\,2180 and corresponding offset field
(top panels), and for NGC\,3680 (bottom panels).}
\label{fig3}
\end{figure}

In Fig.~\ref{fig3} we show the $\jj\times\jh$ CMDs for each cluster (left panels)
along with the corresponding (same area) offset fields (right panels). In order to 
maximize the cluster/background contrast, the CMDs have been built with stars extracted 
within 10\arcmin\ and 15\arcmin, respectively for NGC\,2180 and NGC\,3680. According to 
the analysis in Sect.~\ref{StructAnal}, these dimensions are smaller than the limiting
radius of each cluster.

NGC\,2180 can be recognized as a cluster by the presence of the MS and a group
of bright giants. These features are not present in its comparison field (top-right panel). 
Since the 6 giant stars are located within the central 10\arcmin, we recalculated the central
coordinates of NGC\,2180 as the average values for these stars (Sect.~\ref{targets}). 

The MS in the CMD of NGC\,3680 is well-defined, particularly near the turnoff, and
the presence of giants is also clear.

In all CMDs shown in Fig.~\ref{fig3}, the contribution of disk stars is seen as nearly
vertical sequences at $\jh\approx0.6$ and $\jh\approx0.35$. These contributions, as well 
as that of faint and spurious detections, has to be properly taken into account in order 
to isolate the cluster members. Indeed, it is interesting to note that the number of sources 
in the CMD in the direction of NGC\,2180 is 1410, while in the CMD of its comparison field, 
this number is 1399. For NGC\,3680, the above numbers are 1887 and 1890, respectively for 
cluster and comparison field.

\section{Cluster structure}
\label{StructAnal}

\begin{figure} 
\resizebox{\hsize}{!}{\includegraphics{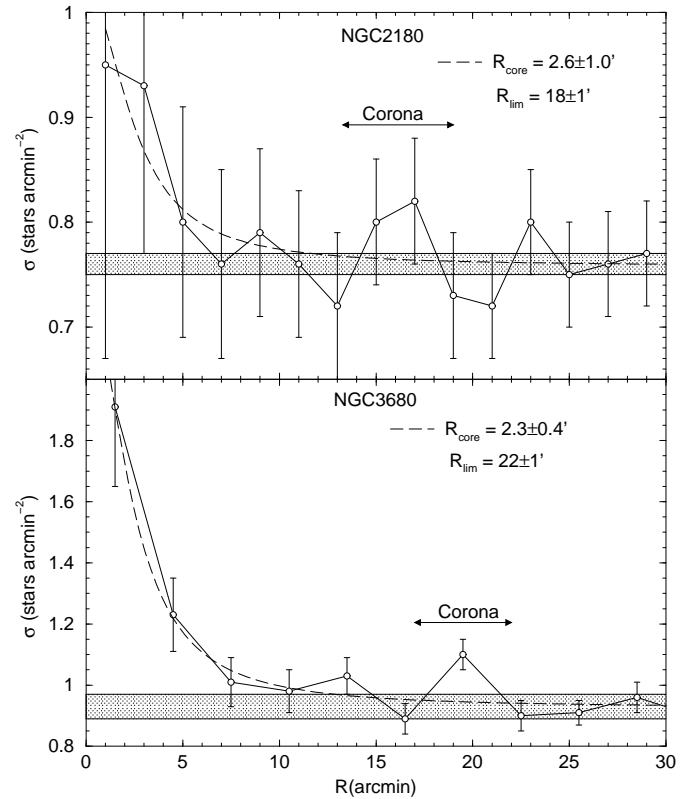}}
\caption[]{Radial distribution of surface star density. The average background levels
are shown as shaded rectangles;  $1-\sigma$ Poissonian errors are also shown. For 
both clusters, magnitude ($\jj<15.0$) and colour cutoffs have been applied to the object 
and offset fields. The dashed lines show a surface density profile  (King law) fit 
to the radial distribution of stars; the resulting core radius for each cluster is indicated.}
\label{fig4}
\end{figure}

The overall cluster structure is analysed by means of the star density radial 
distribution, defined as the number of stars per area in the direction of a cluster,
which is shown in Fig.~\ref{fig4} for NGC\,2180 (top panel) and NGC\,3680 (bottom 
panel). 

Before counting stars, we applied a cutoff ($\jj<15.0$) to both clusters
and corresponding offset fields to avoid oversampling, i.e. to avoid spatial 
variations in the number of faint stars which are numerous, affected by large errors, 
and may include spurious detections. Colour filters in the CMDs have 
also been applied to both clusters and corresponding offset fields, in order to account
for the contamination of the Galaxy. This procedure has been applied in the 
analysis of the open cluster M\,67 (Bonatto \& Bica \cite{BB2003}). As a result of the 
filtering process, the number of stars in the direction of NGC\,2180 turns out to be 
253, compared to 238 in its offset field. For NGC\,3680, these numbers are 755 and 668, 
respectively for cluster and comparison field. The radial distributions have been determined 
by counting stars inside concentric annuli with a step of 2.0\arcmin\ in radius for
NGC\,2180 and 2.5\arcmin\ for NGC\,3680,  and dividing the number of stars by the
respective annulus area. The background contribution level, shown in 
Fig.~\ref{fig4} as shaded rectangles, corresponds to the average number of 
stars present in the external annuli.

The radial density profile of NGC\,2180 (top panel) is not smooth, presenting bumps 
and dips with respect to the background level. The relatively small number of stars
in this cluster is reflected  by the large  $1-\sigma$ Poissonian error bars. Even so, NGC\,2180
still presents a central concentration of stars for $R<5\arcmin$ as well as an excess
in the corona at $14\arcmin\le R\le18\arcmin$. Considering the profile fluctuations with
respect to the background level, most of the cluster's stars can be considered to be contained
within a radius of $\approx10\arcmin$. However, a corona (Sect.~\ref{Fund_par}) is 
detected, and we adopt as limiting radius $\rlim\approx18\arcmin$. The limiting radius
corresponds to the radius at which the cluster's profile merges into the background 
level (Fig.~\ref{fig4}). On the other hand, the radial density profile
of NGC\,3680 (bottom panel) is smooth and presents a well-defined central concentration
of stars, as well as an excess in the corona at $17\arcmin\le R\le22\arcmin$. Its limiting
radius lies at $\rlim\approx22\arcmin$.

Although the spatial shape of the clusters may not be perfectly spherical,  King law 
(\cite{King1966}) is still useful to derive first order structural parameters. A cluster core 
radius \rc\ can be calculated by fitting a  King surface density profile $\sigma(R)=
\frac{\sigma_0}{1+\left(R/R_{core}\right)^2}$\ to the background-subtracted radial distribution 
of stars.  The fits have been performed using a non-linear least-squares fit routine which 
uses the error bars as weights. The resulting fits are shown in Fig.~\ref{fig4}, as dashed 
lines. The radial density profile of NGC\,3680 (bottom panel) follows well  King law, with 
a resulting core radius  $\rc=2.3\pm0.4\arcmin$.  The large $1-\sigma$ Poissonian errors 
and non-uniform density profile of NGC\,2180 produce a significant uncertainty in the resulting 
core radius, $\rc=2.6\pm1.0\arcmin$. Using the cluster distances derived in Sect.~\ref{Fund_par} 
below, the linear core radii turn out to be $\rc=0.7\pm0.3$\,pc and $\rc=0.7\pm0.1$\,pc, 
respectively for NGC\,2180 and NGC\,3680. Finally, the angular diameters ($2\times\rlim$) 
of $36\pm2\arcmin$ and $44\pm2\arcmin$, convert to linear limiting diameters of $9.5\pm1.2$\,pc 
and $12.8\pm1.3$\,pc, respectively for NGC\,2180 and NGC\,3680. 

\section{Fundamental parameters}
\label{Fund_par}

To maximize cluster membership probability, the analyses in the following two sections 
will be restricted to stars extracted within \rlim\ (Sect.~\ref{StructAnal}). 
Cluster parameters will be derived using solar metallicity Padova isochrones from 
Girardi et al. (\cite{Girardi2002}) computed with the 2MASS J, H and K$_S$ filters 
(available at {\em http://pleiadi.pd.astro.it}).
The solar metallicity isochrones have been selected to be consistent with the results
of Eggen (\cite{Eggen1969}) and Nissen (\cite{Nissen1988}), at least for NGC\,3680.
The 2MASS transmission filters produced isochrones very similar to the Johnson ones, with
differences of at most 0.01 in \jh\ (Bonatto et al. \cite{BBG2004}). For reddening 
and absorption transformations we use R$_V$ = 3.2, and the relations A$_J = 0.276\times$A$_V$
and $\ejh=0.33\times\ebv$, according to Dutra et al. (\cite{DSB2002}) and references
therein. 

\begin{figure}
\resizebox{\hsize}{!}{\includegraphics{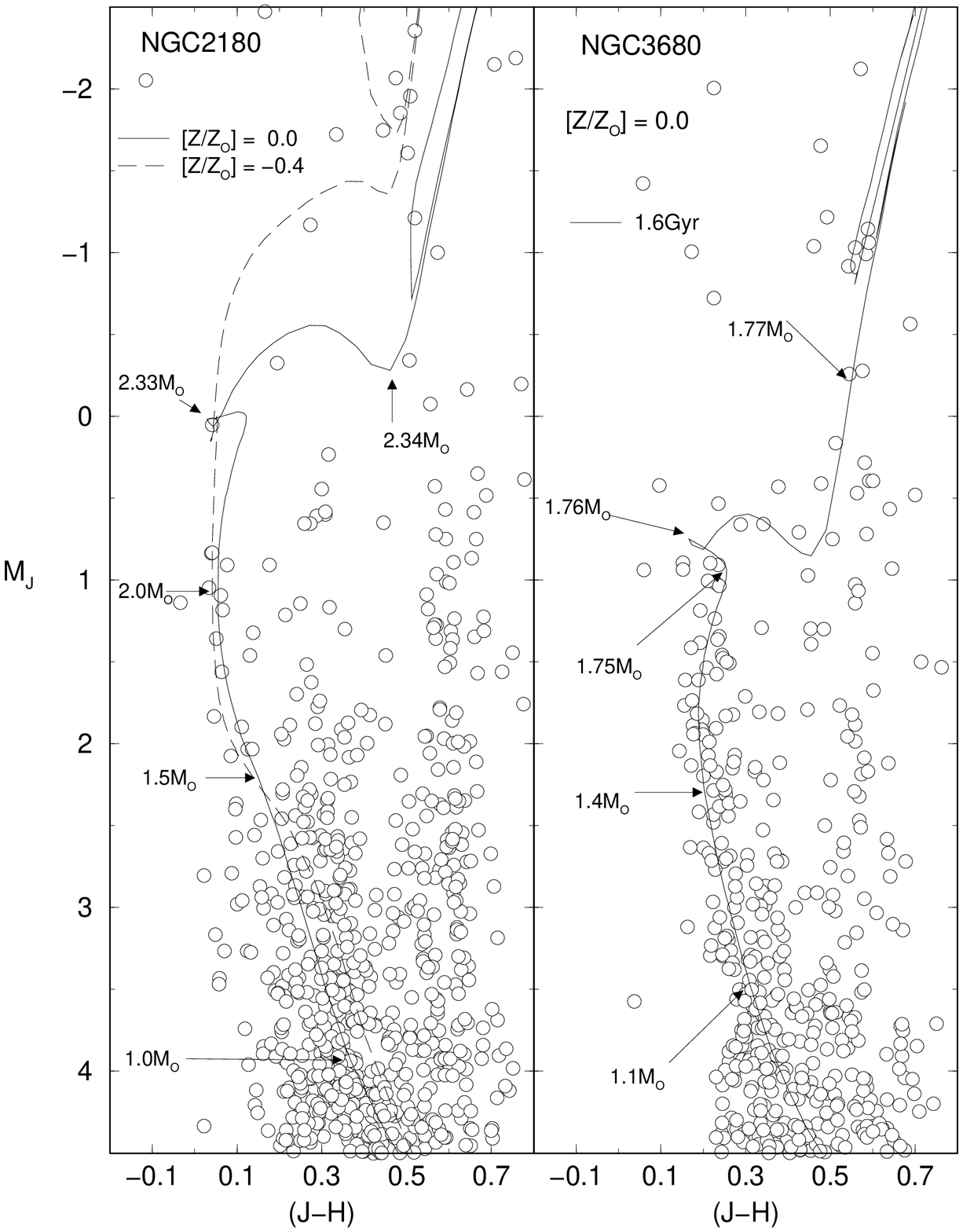}}
\caption[]{Isochrone fits to the $\mj\times\jh$ CMDs of NGC\,2180 (left panel) and NGC\,3680 
(right panel). For NGC\,2180, equally good fits are obtained with the 710\,Myr isochrones with 
metallicity $\zz=0.0$ (solid line) and $\zz=-0.4$ (dashed line). For the 
$\zz=0.0$ solution, $\mM=10.10\pm0.20$, $\ebv=0.0$ and $\ds=1.05\pm0.08$\,kpc, while the
$\zz=-0.4$ solution results in $\mM=9.40\pm0.20$, $\ebv=0.18$ and $\ds=0.76\pm0.06$\,kpc. 
For NGC\,3680 we derive an age of $\age\sim1.6$\,Gyr, $\mM=10.00\pm0.20$, $\ebv=0.00$ and 
$\ds=1.00\pm0.09$\,kpc. Representative stellar masses are indicated for both clusters.}
\label{fig5}
\end{figure}

We show in Fig.~\ref{fig5} the isochrone fits to the $\mj\times\jh$ CMD of NGC\,2180 
(left panel) and NGC\,3680 (right panel). \mj\ values are obtained after applying the 
distance modulus derived below for each cluster. 

The upper MS of NGC\,2180 is relatively depleted of stars, with  a single star near the 
turnoff at $\mj\approx0.0$.  Thus, the solar-metallicity fit which best matches the 
MS features has been obtained with the 710\,Myr isochrone. This solution is shown as a solid 
line in Fig.~\ref{fig5}. The isochrone fit and related uncertainties result in a distance 
modulus $\mM=10.10\pm0.20$, $\ebv=0.0$ and a distance to the Sun $\ds=1.05\pm0.08$\,kpc.  
However, the above isochrone solution fails to reproduce the 6 giants at $\mj\approx-2$
and $\jh\approx0.5$, which might be accounted for by differences in metallicity. Accordingly, 
we present an alternative fit with the subsolar metallicity ($\zz=-0.38$), 710\,Myr isochrone,
which results in a distance modulus $\mM=9.40\pm0.20$, $\ebv=0.18$ and $\ds=0.76\pm0.06$\,kpc. 
According to this solution (dashed line), the 6 bright stars would be, in fact, the giant clump 
of NGC\,2180. The giant clump in NGC\,2180 is similar to that of the Hyades cluster (e.g. Perryman 
et al. \cite{Perryman1998}). Both clusters have similar ages. Since  the CMD features below 
the turnoff are equally well reproduced by both isochrones, the intrinsic metallicity of NGC\,2180 
is probably in the range $-0.4\leq\zz\leq0.0$. In the same way, we adopt as distance to the Sun 
the average of the values derived for both isochrones, i.e. $\ds=0.91\pm0.15$\,kpc. With the above 
distance to the Sun, the galactocentric distance of NGC\,2180 becomes $8.8\pm0.1$\,kpc.  In 
the subsequent calculations we will adopt the solar metallicity solution as reference, 
particularly to derive the LFs and MFs.

The presence of a well-defined turnoff and giant stars in the CMD of NGC\,3680 
constrain the isochrone fits to the $\age=1.6$\,Gyr solution. For this cluster
we derive $\mM=10.00\pm0.20$, $\ebv=0.00$ and $\ds=1.00\pm0.09$\,kpc. The galactocentric 
distance of NGC\,3680 is $7.8\pm0.1$\,kpc. The present
age estimate, using 2MASS photometry, is in close agreement with those derived by
Kozhurina-Platais et al. (\cite{Kozhu1997}) and Bruntt et al. (\cite{Bruntt1999}).

\section{Luminosity and mass functions}
\label{LumFunc}

In this section we analyze the observed star counts as a function of magnitude 
(LF) and mass (MF) as well as their spatial dependence.

The accurate determination of a cluster's LF (or MF) suffers from some problems, in 
particular {\it (i)} the contamination of cluster members by field stars, {\it (ii)} 
the observed incompleteness at low-luminosity (or low-mass) stars, and  {\it (iii)} 
the mass segregation, which may affect even poorly populated, relatively young
clusters (Scalo \cite{Scalo1998}). The 2MASS uniform sky coverage allows one to overcome, 
at least in part, points {\it (i)} --- since suitable offset fields can be selected
around the cluster and {\it (iii)} --- the entire cluster area can be included in 
the analyses. Thus, advanced stages of mass segregation would affect more significantly
the analysis of very old, dynamically evolved clusters (e.g. M\,67, Bonatto \& Bica 
\cite{BB2003}). 

\begin{figure}
\resizebox{\hsize}{!}{\includegraphics{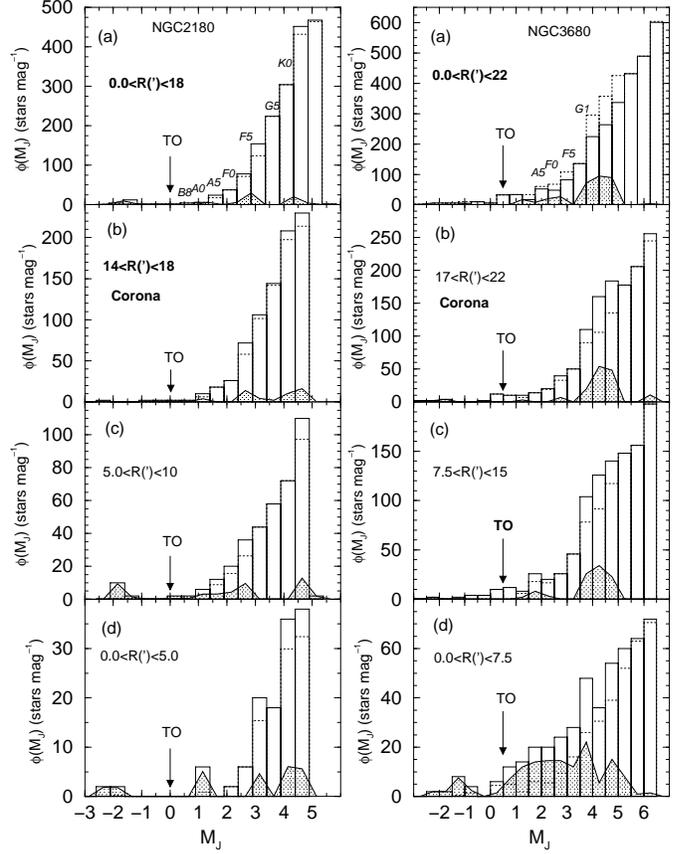}}
\caption[]{Luminosity functions ($\phi(\mj)$) in terms of the absolute magnitude \mj. 
Magnitude ($\jj<15.0$) and colour cutoffs have been applied to the object and offset 
fields. Continuous line: star counts in the cluster's area; dotted line: star counts 
in the offset field; shaded area: background-subtracted LF. The turnoff (TO) is indicated 
in each panel.}
\label{fig6}
\end{figure}

In Fig.~\ref{fig6} we show the LFs ($\phi(\mj)$) in the J filter (shaded area) for 
the two clusters, built as the difference of the number of stars in a given magnitude 
bin between object (continuous line) and average offset field (dotted line). The LFs 
are given in terms of the absolute magnitude \mj, obtained after applying the distance 
modulus derived in Sect.~\ref{Fund_par} for each cluster, and the bin in magnitude used
is $\Delta\mj=0.50$\,mag. We remind that the LFs in Fig.~\ref{fig6} are built after 
applying magnitude ($\jj<15.0$) and colour cutoffs to the objects and their offset 
fields (Sect.~\ref{Fund_par}). The LFs for different spatial regions of NGC\,2180 
are in the left panels, and those of NGC\,3680 are in the right panels.

To search for spatial variations in the stellar content, we built LFs for different
regions in and around the clusters, according to the structures present in the radial
density profiles (Fig.~\ref{fig4}). Thus, the first LF encompasses the  central
region of each cluster (panels (d)), $0.0\arcmin\le R\le 5.0\arcmin$ and $0.0\arcmin\le 
R\le 7.5\arcmin$, respectively for NGC\,2180 and NGC\,3680; the second one corresponds 
to the outskirts (panels (c)), $5.0\arcmin\le R\le 10.0\arcmin$ and $7.5\arcmin\le R\le
15.0\arcmin$, and the third LF to the corona (panels (b)).
In panels (a) we show  the overall LFs ($0.0\arcmin\le R\le\rlim$). The background 
$\phi(\mj)$ has been scaled to match the projected area of each region. 
Representative MS spectral types (adapted from Binney \& Merrifield \cite{Binney1998}) 
are shown in the top panels. The turnoff is indicated in all panels. Giants are contained 
within 10\arcmin\ in NGC\,2180 and 15\arcmin\ in NGC\,3680.

The severe depletion of MS stars in NGC\,2180 can be clearly seen in the internal 
background-subtracted LFs (Fig.~\ref{fig6}, left panels (c) and (d)) which, in fact, 
are similar to each other. The above spatial properties of the LFs probably reflect 
mass segregation followed by advanced and significant Galactic tidal stripping effects 
(severe depletion of stars in the corona). Indeed, the overall cluster and offset field 
LFs (left panel (a)) are very similar, indicating that the original stellar content of 
NGC\,2180 is already nearly dispersed into the background population.  In order to
reach such an advanced dynamical state in a time span of $\sim700$\,Myr it is reasonable
to assume that the initial cluster was not massive.

On the other hand, the central LF of NGC\,3680 (Fig.~\ref{fig6}, right panel (d)) is 
nearly flat from the turnoff (spectral type $\sim$\,A\,0) to the turnover ($\sim$\,G\,1), 
confirming the central depletion of lower-MS stars found by Anthony-Twarog et al. 
(\cite{Twa1997}). This effect may be accounted for by mass segregation alone. Indeed,
the $7.5\arcmin\le R\le15\arcmin$ and corona regions (right panels (c) and (b)) are 
still well populated by low-mass ($\sim$\,G\,0) stars, suggesting that the Galactic 
tidal stripping has not yet been effective in severely depleting this cluster. In 
this sense, NGC\,3680 is very similar to the $\age\approx3.2$\,Gyr, mass-segregated 
open cluster M\,67, as spatially analysed in Bonatto \& Bica (\cite{BB2003}).

\begin{figure}
\resizebox{\hsize}{!}{\includegraphics{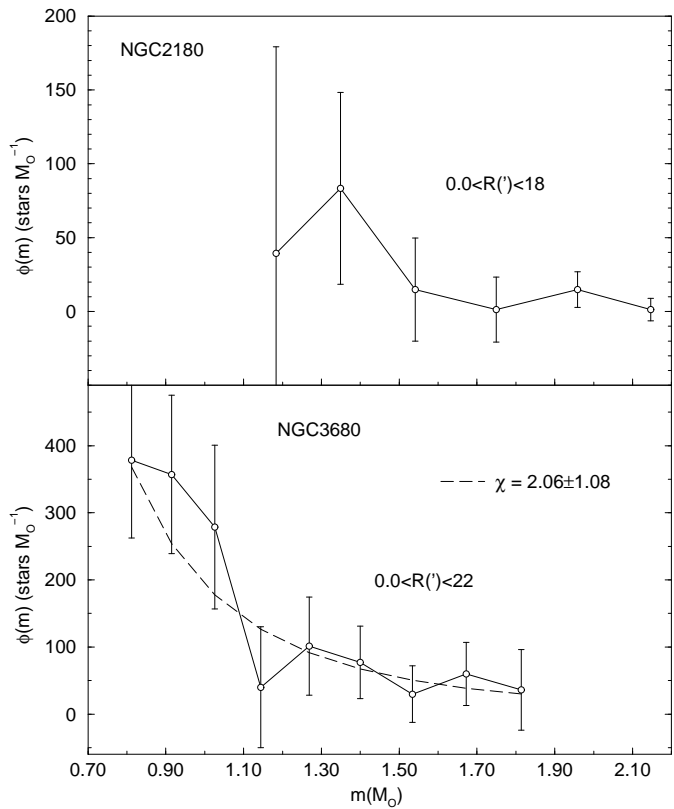}}
\caption[]{Turnoff--turnover mass functions. The MF fit $\phi(m)\propto m^{-(1+\chi)}$ 
for NGC\,3680 is shown as a dashed line.}
\label{fig7}
\end{figure}

The overall LFs in Fig.~\ref{fig6} (top panels), restricted to the region between the 
turnoff and turnover, have been converted to MFs according to $\phi(m)=
\phi(\mj)\left|\frac{dm}{d\mj}\right|^{-1}$. We used the stellar mass-luminosity relations
taken from the Padova isochrones (Sect.~\ref{Fund_par}), 710\,Myr 
for NGC\,2180 and 1.6\,Gyr for NGC\,3680. Since we are interested in obtaining only an estimate
of the mass in NGC\,2180, we used the solar metallicity isochrone to derive the mass--luminosity 
relation. The resulting MFs,  including $1-\sigma$ error bars, are shown in Fig.~\ref{fig7}.

Despite the non-uniform overall MF of NGC\,3680, we applied a fit of $\phi(m)\propto m^{-(1+\chi)}$
to the turnoff ($\approx1.8\,\ms$, $\approx$\,A\,0) --- turnover ($\approx1.0\,\ms$, $\approx$\,G\,1)
region, which resulted in a slope $\chi=2.06\pm1.08$, a steeper MS than a Salpeter's
$\chi=1.35$. Numerical integration of the MF in the
above mass interval, taking into account the uncertainties, resulted in a MS mass of 
$\mMs=111\pm24\,\ms$. We estimate the mass stored in the giants (\mg) by counting the 
number of stars brighter than the turnoff present in the overall LFs, and multiplying this number 
by the  mass at the turnoff, resulting in $\mg\approx19\,\ms$ for NGC\,3680. Thus, the present 
observed stellar mass in NGC\,3680 is $\mobs=130\pm24\,\ms$. 

Assuming that the very low-mass stars have not yet been stripped away from the cluster,
the total mass locked up in stars can be estimated by directly extrapolating the present MF 
fit down to the theoretical stellar low-mass end $m_{low}=0.08\,\ms$. The resulting total
stellar mass in NGC\,3680 is $\sim(2.4\pm1.2)\times10^3\,\ms$ which, within errors, is similar 
to the value found by Nordstr\"om et al. (\cite{Nordstrom1996}). This agreement is interesting, 
considering the different methods of taking into account the stellar background. On the other 
hand, Kroupa, Tout \& Gilmore (\cite{KTG1991}) and Kroupa (\cite{Kroupa2001}) presented evidence
that the MFs of most globular and open clusters flatten below $\sim0.5\,\ms$. As a consequence, 
the total stellar mass in NGC\,3680 
would be less than the value derived above, since most of the stars in a cluster are expected
to be found in the low-mass range. Accordingly, we derive a more conservative total mass value 
for NGC\,3680 assuming the universal IMF of Kroupa (\cite{Kroupa2001}), in which $\chi=0.3\pm0.5$ 
for $0.08\,\ms - 0.50\,\ms$, and $\chi=1.3\pm0.3$ for $0.50\,\ms - 0.80\,\ms$. For $0.80\,\ms - 
1.80\,\ms$ we adopt the value derived in this paper, $\chi=2.06\pm1.08$. As expected, the 
resulting total stellar mass decreases to $546\pm206\,\ms$. 

Similar to NGC\,3680, we derive a giant mass of $\mg\approx16\,\ms$ for NGC\,2180.
However, the considerable variations in the MF do not allow a statistically significant 
fit. Consequently, the present observed stellar mass in NGC\,2180 has been estimated by 
numerically integrating its MF (Fig.~\ref{fig7}, top panel) and adding \mg\ to this value. 
The resulting observed stellar mass is $\mobs\sim47\pm7\,\ms$. Considering a severe
low-mass star depletion (Bergond et al. \cite{Bergond2001}, de la Fuente Marcos 
\cite{delaF1996}) for NGC\,2180, it would certainly be less massive than NGC\,3680,
which is older and at a somewhat smaller Galactocentric distance  (Sect.~\ref{Comp}).

Taking into account the mass estimates, the derived ages of both clusters are
more than an order of magnitude larger than the corresponding $t_{relax}$, which 
is consistent with the significant presence of mass-segregation effects.

\section{Comparison with other dynamical states}
\label{Comp}

\begin{figure} 
\resizebox{\hsize}{!}{\includegraphics{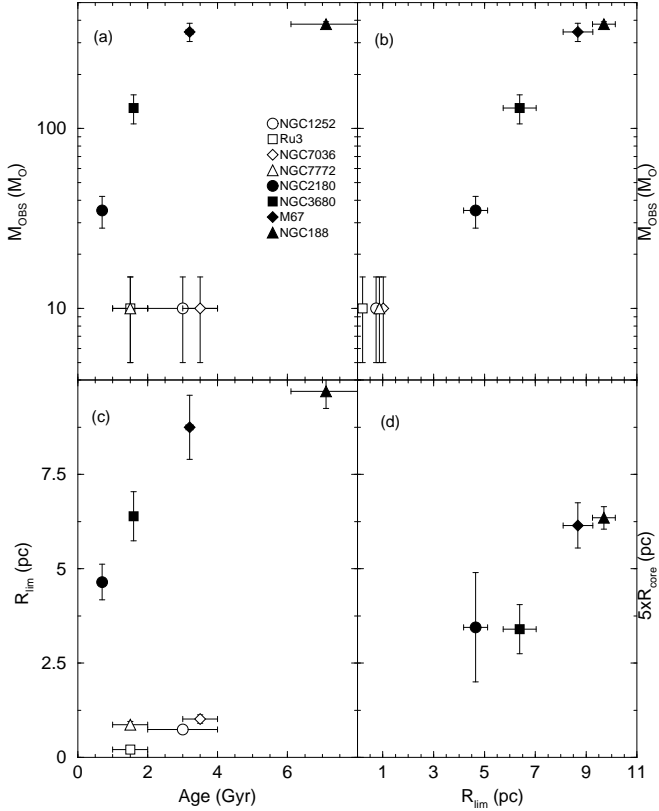}}
\caption[]{NGC\,2180 and NGC\,3680 compared to dynamically evolved open clusters and 
known open cluster remnants. For visualization purposes, the ordinate in panel (d)
corresponds to $5\times\rc$.}
\label{fig8}
\end{figure}

At this point, it may be useful to see where NGC\,2180 and NGC\,3680 fit in the context of 
{\it (i)} well-known dynamically evolved open clusters, and {\it (ii)} poorly populated 
remnants of open clusters. In particular, we will pay attention to parameters intimately 
associated to dynamical evolution, i.e., age, core and limiting radii and stellar mass.
The remnants are NGC\,1252 (Pavani et al. \cite{Pavani2002}), Ruprecht\,3 (Pavani et al. 
\cite{Pavani2003}), NGC\,7036 and NGC\,7772 (Carraro \cite{Carraro2002}). The old open 
clusters are M\,67 (Bonatto \& Bica \cite{BB2003}) and NGC\,188 (Bonatto, Bica \& Santos 
\cite{BBJF2004}), both more massive than NGC\,2180 and NGC\,3680. All these objects 
have been analysed by means of 2MASS photometry following the same methods as in the case 
of NGC\,2180 and NGC\,3680, thus ensuring homogeneity in terms of analysis and derived
parameters. Relevant parameters for the objects are given in Table~\ref{tab1}. The results 
are shown in Fig.~\ref{fig8}. Part of the parameters for NGC\,1252, Ruprecht\,3, NGC\,7036 
and NGC\,7772 have been derived using 2MASS data, following a similar analysis as that used 
for NGC\,2180 and NGC\,3680.

As expected, the structural parameters of the intrinsic open clusters present
a strong dependence with time. This is particularly true for the limiting radius 
(panel (c)), which increases linearly with time, probably as a consequence of the Galactic 
tidal pull. The observed mass, on the other hand, initially increases linearly with time 
and then reaches a saturation threshold (panel (a)). This threshold is defined by
M\,67 and NGC\,188, both dynamically advanced open clusters, more massive than NGC\,2180 
and NGC\,3680,  with strong effects of mass segregation. The relations of the core radius 
to the limiting radius (panel (d)),
and of \rlim\ to \mobs\ (panel (b)), present a similar pattern, linear increase first
followed by a saturation. This behaviour may be accounted for by a combination of the 
Galactic tidal pull (increasing \rlim\ with time) and mass segregation (initial core 
formation and growth, followed by core stabilization (de la Fuente Marcos \cite{delaF1996})). 
In the far future, these massive open clusters will leave behind only a core, with
most of the low-mass stars dispersed into the background (de la Fuente Marcos 
\cite{delaF1996}). 

\begin{table}
\caption[]{Open cluster and remnant parameters.}
\begin{tiny}
\label{tab1}
\renewcommand{\tabcolsep}{0.60mm}
\begin{tabular}{lccccccccc}
\hline\hline
Cluster&$\ell$&$b$ & d$_{GC}$&Age & \rc & \rlim & \mobs \\
       & ($^\circ$) & ($^\circ$)&(kpc) &(Gyr)& (pc)& (pc)  &(\ms) \\
\hline
NGC\,2180&203.85&-7.01&9.0$^a$&$0.7\pm0.1^a$& $0.7\pm0.3^a$ & $4.7\pm0.6^a$ & $47\pm7^a$\\
NGC\,3680&286.76&16.92&8.8$^a$&$1.6\pm0.1^a$& $0.7\pm0.1^a$ & $6.4\pm0.6^a$ & $130\pm24^a$\\
M\,67&215.68&31.93&8.6$^b$&$3.2\pm0.1^b$& $1.2\pm0.1^b$ & $8.7\pm0.8^b$ & $344\pm40^b$ \\
NGC\,188&122.85&22.39&8.9$^c$&$7.1\pm0.1^c$& $1.3\pm0.1^c$ & $9.7\pm0.4^c$ & $380\pm12^c$ \\
NGC\,1252&274.08&-50.83&7.8$^d$&$3.0\pm1.0^d$& --- & $0.7\pm0.1^a$& $10\pm5^a$ \\
NGC\,7772&102.73&-44.27&8.4$^d$&$1.5\pm0.5^e$&--- & $0.9\pm0.1^g$ &$10\pm5^a$  \\
NGC\,7036&64.54&-21.44&7.6$^d$&$3.5\pm0.5^e$&--- &  $1.0\pm0.1^g$ &$10\pm5^a$  \\
Ruprecht\,3&238.77&-14.81&8.4$^d$&$1.5\pm0.5^f$& --- & $0.2\pm0.0^a$ & $10\pm5^a$\\
\hline
\end{tabular}
\end{tiny}
\begin{list}{Table Notes.}
\item Column (4): Galactocentric distance; Data sources are: (a) - this paper, using 2MASS; (b) - Bonatto \& Bica (\cite{BB2003}); 
(c) - Bonatto, Bica \& Santos (\cite{BBJF2004}) ; (d) - Pavani et al. (\cite{Pavani2002}); 
(e) - Carraro (\cite{Carraro2002}); (f) - Pavani et al. (\cite{Pavani2003}); (g) - this paper, 
using Carraro (\cite{Carraro2002}). 
\end{list}
\end{table}

With respect to the structural parameters, the open clusters seem to follow a well-defined 
sequence (panels (a) -- (d)). The remnants, however, follow this sequence only for
$\rlim\times\mobs$ (panel (b)), since most of their stellar content  may have already 
been dispersed into the background.

NGC\,2180 is consistently closer than NGC\,3680 to the loci occupied by the remnants
in panels (a), (b) and (c). Despite being younger than NGC\,3680, the less-massive
nature of NGC\,2180 probably accelerated its evolutionary time-scale, setting it in 
a more dynamically advanced state than NGC\,3680. 

Interestingly, the remnants dealt with in this paper have Galactocentric distances
smaller that those of the evolved open clusters (Column~4 of Table~\ref{tab1}). Again, 
this suggests the effects of the Galactic tidal pull (external evolution).  However,
to draw objective conclusions with respect to open cluster evolution it is necessary
to take separately into account selection and intrinsic evolution effects. To further 
explore these issues we intend to analyze a larger sample of clusters which span a variety
of photometric, structural and dynamical properties, using the same methods employed
in the present paper.

\section{Concluding remarks}
\label{Conclu}

As a consequence of the internal dynamical evolution and the relentless
Galactic tidal pull, most open clusters are expected to evaporate completely 
in less than 1\,Gyr, leaving behind a remnant with characteristics which
depend on the cluster's initial conditions. Only the more massive open clusters 
may survive to old ages. In this context, the observation of an actually dissolving 
open cluster becomes very interesting to check existing theories on dynamical
evolution of stellar systems, N-body codes in particular, as well as to test
stellar evolution theories.

In the present work we analyse the physical structure, stellar content and dynamical
state of the overlooked open cluster NGC\,2180. We also examine in detail the
intermediate-age open cluster NGC\,3680, formerly considered to be in the last
stages of its dynamical evolution. The present analyses make use mostly of J and H 
2MASS All Sky data release photometry.

NGC\,2180 presents a non-uniform radial distribution of stars (Sect.~\ref{StructAnal}),
with significant  $1-\sigma$ Poissonian error bars due to the small number of member 
stars. Its radial density profile has a central concentration of stars, as well as a corona 
(Fig.~\ref{fig4}). From a  King law fit we estimate $\rc=0.7\pm0.3$\,pc and a 
linear limiting diameter of $9.5\pm1.2$\,pc. Its $\mj\times\jh$ CMD (Fig.~\ref{fig5}) is 
depleted of stars near the turnoff, and can be fitted with 710\,Myr isochrones of solar 
and sub-solar metallicity. The $\zz=0.0$ solution results in $\mM=10.10\pm0.20$, $\ebv=0.0$ 
and a distance to the Sun $\ds=1.05\pm0.08$\,kpc, while the $\zz=-0.4$ solution gives 
$\mM=9.40\pm0.20$, $\ebv=0.18$ and $\ds=0.76\pm0.06$\,kpc. Thus, we adopt as distance to 
the Sun $\ds=0.91\pm0.15$\,kpc, which puts NGC\,2180 at a galactocentric distance of
$8.8\pm0.1$\,kpc. Mass segregation and advanced Galactic tidal stripping on NGC\,2180 are 
reflected on the spatial properties of its LFs (Fig.~\ref{fig6}). Low-mass stars are severely 
depleted from the MS in each region, from the center to the cluster's limiting radius. 
In addition, the MS of the corona, although depleted as well, is slightly more populated 
of low-mass stars than the MS of more internal regions. The observed stellar mass in 
NGC\,2180 is $\sim47\pm7\,\ms$.

NGC\,3680 presents a uniform radial distribution of stars, with a well-defined core 
and a corona, with $\rc=0.7\pm0.1$\,pc and a linear limiting diameter of 
$12.8\pm1.3$\,pc (Sect.~\ref{StructAnal}). Its $\mj\times\jh$ CMD (Fig.~\ref{fig5}) 
presents a nearly complete MS, including the turnoff and giants. We derive an 
age of $\age\approx1.6$\,Gyr, $\mM=10.00\pm0.20$, $\ebv=0.00$ and $\ds=1.00\pm0.09$\,kpc,
in reasonable agreement with previous works. The LF of the central region of NGC\,3680
(Fig.~\ref{fig6}) is depleted of low-mass stars which, contrary to what is observed 
in NGC\,2180, are still present in the external region and corona. Thus, Galactic 
tidal stripping has not yet been effective in severely depleting NGC\,3680 of stars.
For NGC\,3680, a MF fit $\phi(m)\propto m^{-(1+\chi)}$ resulted in a slope
$\chi=2.06\pm1.08$, and in an observed stellar mass (MS and giants) of $\approx130\pm24\,\ms$.  
Extrapolating the MF fit down to the theoretical low-mass end $m_{low}=0.08\,\ms$, 
the total stellar mass in NGC\,3680 turns out to be $\sim(2.4\pm1.2)\times10^3\,\ms$, which 
agrees with previous estimates, within uncertainties.  Assuming a more representative
IMF (Kroupa \cite{Kroupa2001}), which flattens for masses below $\sim0.5\,\ms$, the 
total stellar mass in NGC\,3680 turns out to be $546\pm206\,\ms$. 

Finally, comparing NGC\,2180 and NGC\,3680 with clusters in other dynamical states,
{as well as open cluster remnants}, we found that the less-massive nature of NGC\,2180 
put it closer to cluster remnants than NGC\,3680.

The above arguments lead us to conclude that NGC\,2180 is in a more advanced 
dynamical state than NGC\,3680, on its way to become a fossil cluster. Thus,
NGC\,2180 may be the missing link between evolved open clusters and remnants.

\begin{acknowledgements}
We thank the referee, Dr. R. de Grijs, for interesting remarks.
This publication makes use of data products from the Two Micron All Sky Survey, which is a joint 
project of the University of Massachusetts and the Infrared Processing and Analysis Center/California 
Institute of Technology, funded by the National Aeronautics and Space Administration and the National 
Science Foundation. We employed catalogues from CDS/Simbad (Strasbourg) and Digitized Sky Survey 
images from the Space Telescope Science Institute (U.S. Government grant NAG W-2166) obtained using 
the extraction tool from CADC (Canada). We also made use of the WEBDA open cluster database. We 
acknowledge support from the Brazilian Institution CNPq.
\end{acknowledgements}

%

\end{document}